\documentstyle[12pt]{article}

\catcode`\@=11
\long\def\@makefntext#1{ %\parindent 1em
\protect\noindent \hbox to 3.2pt {\hskip-.9pt
$^{{\ninerm\@thefnmark}}$\hfil}#1\hfill} %can be used

\def\thefootnote{\fnsymbol{footnote}}
 \def\@makefnmark{\hbox to 0pt{$^{\@thefnmark}$\hss}}  %original

\def\ps@myheadings{\let\@mkboth\@gobbletwo
\def\@oddhead{\hbox{} %\sl
\rightmark\hfil\ninerm\thepage}
\def\@oddfoot{}\def\@evenhead{\ninerm\thepage\hfil %\sl
\leftmark\hbox{}}\def\@evenfoot{}
\def\sectionmark##1{}\def\subsectionmark##1{}}

\begin{document}

%----------------------------PROCSLA.STY---------------------------------------
\newcommand{\symbolfootnote}{\renewcommand{\thefootnote}
	{\fnsymbol{footnote}}}
\renewcommand{\thefootnote}{\fnsymbol{footnote}}
\newcommand{\alphfootnote}
	{\setcounter{footnote}{0}
	 \renewcommand{\thefootnote}{\sevenrm\alph{footnote}}}

%------------------------------------------------------------------------------
%NEW DEFINED SECTION COMMANDS
\newcounter{sectionc}\newcounter{subsectionc}\newcounter{subsubsectionc}
\renewcommand{\section}[1] {\vspace{0.6cm}\addtocounter{sectionc}{1}
\setcounter{subsectionc}{0}\setcounter{subsubsectionc}{0}\noindent
	{\bf\thesectionc. #1}\par\vspace{0.4cm}}
\renewcommand{\subsection}[1] {\vspace{0.6cm}\addtocounter{subsectionc}{1}
	\setcounter{subsubsectionc}{0}\noindent
	{\it\thesectionc.\thesubsectionc. #1}\par\vspace{0.4cm}}
\renewcommand{\subsubsection}[1]
{\vspace{0.6cm}\addtocounter{subsubsectionc}{1}
	\noindent {\rm\thesectionc.\thesubsectionc.\thesubsubsectionc.
	#1}\par\vspace{0.4cm}}
\newcommand{\nonumsection}[1] {\vspace{0.6cm}\noindent{\bf #1}
	\par\vspace{0.4cm}}

%NEW MACRO TO HANDLE APPENDICES
\newcounter{appendixc}
\newcounter{subappendixc}[appendixc]
\newcounter{subsubappendixc}[subappendixc]
\renewcommand{\thesubappendixc}{\Alph{appendixc}.\arabic{subappendixc}}
\renewcommand{\thesubsubappendixc}
	{\Alph{appendixc}.\arabic{subappendixc}.\arabic{subsubappendixc}}

\renewcommand{\appendix}[1] {\vspace{0.6cm}
        \refstepcounter{appendixc}
        \setcounter{figure}{0}
        \setcounter{table}{0}
        \setcounter{equation}{0}
        \renewcommand{\thefigure}{\Alph{appendixc}.\arabic{figure}}
        \renewcommand{\thetable}{\Alph{appendixc}.\arabic{table}}
        \renewcommand{\theappendixc}{\Alph{appendixc}}
        \renewcommand{\theequation}{\Alph{appendixc}.\arabic{equation}}
%       \noindent{\bf Appendix \theappendixc. #1}\par\vspace{0.4cm}}
        \noindent{\bf Appendix \theappendixc #1}\par\vspace{0.4cm}}
\newcommand{\subappendix}[1] {\vspace{0.6cm}
        \refstepcounter{subappendixc}
        \noindent{\bf Appendix \thesubappendixc. #1}\par\vspace{0.4cm}}
\newcommand{\subsubappendix}[1] {\vspace{0.6cm}
        \refstepcounter{subsubappendixc}
        \noindent{\it Appendix \thesubsubappendixc. #1}
	\par\vspace{0.4cm}}

%------------------------------------------------------------------------------
%MARCO FOR ABSTRACT BLOCK
\def\abstracts#1{{
	\centering{\begin{minipage}{30pc}\tenrm\baselineskip=12pt\noindent
	\centerline{\tenrm ABSTRACT}\vspace{0.3cm}
	\parindent=0pt #1
	\end{minipage} }\par}}

%------------------------------------------------------------------------------
%NEW MACRO FOR BIBLIOGRAPHY
\newcommand{\bibit}{\it}
\newcommand{\bibbf}{\bf}
\renewenvironment{thebibliography}[1]
	{\begin{list}{\arabic{enumi}.}
	{\usecounter{enumi}\setlength{\parsep}{0pt}
%1.25cm IS STRICTLY FOR PROCSLA.TEX ONLY
\setlength{\leftmargin 1.25cm}{\rightmargin 0pt}
%0.52cm IS FOR NEW DATA FILES
%\setlength{\leftmargin 0.52cm}{\rightmargin 0pt}
	 \setlength{\itemsep}{0pt} \settowidth
	{\labelwidth}{#1.}\sloppy}}{\end{list}}

%------------------------------------------------------------------------------
%FOLLOWING THREE COMMANDS ARE FOR 'LIST' COMMAND.
\topsep=0in\parsep=0in\itemsep=0in
\parindent=1.5pc

%LIST ENVIRONMENTS
\newcounter{itemlistc}
\newcounter{romanlistc}
\newcounter{alphlistc}
\newcounter{arabiclistc}
\newenvironment{itemlist}
    	{\setcounter{itemlistc}{0}
	 \begin{list}{$\bullet$}
	{\usecounter{itemlistc}
	 \setlength{\parsep}{0pt}
	 \setlength{\itemsep}{0pt}}}{\end{list}}

\newenvironment{romanlist}
	{\setcounter{romanlistc}{0}
	 \begin{list}{$($\roman{romanlistc}$)$}
	{\usecounter{romanlistc}
	 \setlength{\parsep}{0pt}
	 \setlength{\itemsep}{0pt}}}{\end{list}}

\newenvironment{alphlist}
	{\setcounter{alphlistc}{0}
	 \begin{list}{$($\alph{alphlistc}$)$}
	{\usecounter{alphlistc}
	 \setlength{\parsep}{0pt}
	 \setlength{\itemsep}{0pt}}}{\end{list}}

\newenvironment{arabiclist}
	{\setcounter{arabiclistc}{0}
	 \begin{list}{\arabic{arabiclistc}}
	{\usecounter{arabiclistc}
	 \setlength{\parsep}{0pt}
	 \setlength{\itemsep}{0pt}}}{\end{list}}

%------------------------------------------------------------------------------
%FIGURE CAPTION
\newcommand{\fcaption}[1]{
        \refstepcounter{figure}
        \setbox\@tempboxa = \hbox{\tenrm Fig.~\thefigure. #1}
        \ifdim \wd\@tempboxa > 6in
           {\begin{center}
        \parbox{6in}{\tenrm\baselineskip=12pt Fig.~\thefigure. #1 }
            \end{center}}
        \else
             {\begin{center}
             {\tenrm Fig.~\thefigure. #1}
              \end{center}}
        \fi}

%TABLE CAPTION
\newcommand{\tcaption}[1]{
        \refstepcounter{table}
        \setbox\@tempboxa = \hbox{\tenrm Table~\thetable. #1}
        \ifdim \wd\@tempboxa > 6in
           {\begin{center}
        \parbox{6in}{\tenrm\baselineskip=12pt Table~\thetable. #1 }
            \end{center}}
        \else
             {\begin{center}
             {\tenrm Table~\thetable. #1}
              \end{center}}
        \fi}

%------------------------------------------------------------------------------
%ACKNOWLEDGEMENT: this portion is from John Hershberger
\def\@citex[#1]#2{\if@filesw\immediate\write\@auxout
	{\string\citation{#2}}\fi
\def\@citea{}\@cite{\@for\@citeb:=#2\do
	{\@citea\def\@citea{,}\@ifundefined
	{b@\@citeb}{{\bf ?}\@warning
	{Citation `\@citeb' on page \thepage \space undefined}}
	{\csname b@\@citeb\endcsname}}}{#1}}

\newif\if@cghi
\def\cite{\@cghitrue\@ifnextchar [{\@tempswatrue
	\@citex}{\@tempswafalse\@citex[]}}
\def\citelow{\@cghifalse\@ifnextchar [{\@tempswatrue
	\@citex}{\@tempswafalse\@citex[]}}
\def\@cite#1#2{{$\null^{#1}$\if@tempswa\typeout
	{IJCGA warning: optional citation argument
	ignored: `#2'} \fi}}
\newcommand{\citeup}{\cite}

%------------------------------------------------------------------------------
%FOR FNSYMBOL FOOTNOTE AND ALPH{FOOTNOTE}
\def\fnm#1{$^{\mbox{\scriptsize #1}}$}
\def\fnt#1#2{\footnotetext{\kern-.3em
	{$^{\mbox{\sevenrm #1}}$}{#2}}}

%------------------------------------------------------------------------------
\font\twelvebf=cmbx10 scaled\magstep 1
\font\twelverm=cmr10 scaled\magstep 1
\font\twelveit=cmti10 scaled\magstep 1
\font\elevenbfit=cmbxti10 scaled\magstephalf
\font\elevenbf=cmbx10 scaled\magstephalf
\font\elevenrm=cmr10 scaled\magstephalf
\font\elevenit=cmti10 scaled\magstephalf
\font\bfit=cmbxti10
\font\tenbf=cmbx10
\font\tenrm=cmr10
\font\tenit=cmti10
\font\ninebf=cmbx9
\font\ninerm=cmr9
\font\nineit=cmti9
\font\eightbf=cmbx8
\font\eightrm=cmr8
\font\eightit=cmti8

% *********************Start of T.S.E. definitions***********************

\newcommand\ba{\begin{array}}
\newcommand\ea{\end{array}}
\newcommand\ben{\begin{equation}}
\newcommand\een{\end{equation}}
\newcommand\beq{\begin{equation}}
\newcommand\eeq{\end{equation}}
\newcommand\bea{\begin{eqnarray}}
\newcommand\eea{\end{eqnarray}}

\newcommand{\sinc}{{\rm sinc}}

% *********************End of T.S.E. definitions***********************

%----------------------START OF DATA FILE------------------------------
\begin{flushright}
Imperial/TP/94-95/47\\
{\tt hep-ph/9504?}\\
{\LaTeX}-ed on \today\\
\end{flushright}
\centerline{\tenbf VORTEX PRODUCTION AT PHASE TRANSITIONS }
\centerline{\tenbf IN NONRELATIVISTIC AND RELATIVISTIC MEDIA }
\baselineskip=16pt
%\centerline{\ninerm (For 20\% Reduction to 6 in. $\times$ 8.5 in. Trim Size)}
\vspace{0.8cm}
\centerline{\tenrm R.J. RIVERS
\footnote{Invited talk at the 3me. Colloque Cosmologie, Paris, June,
1995. E-mail; R.Rivers@IC.AC.UK}}
\baselineskip=13pt
\centerline{\tenit Blackett Lab., Imperial College, Prince Consort Road}
\baselineskip=12pt
\centerline{\tenit London, SW7 2BZ, U.K.}
%\vspace{0.3cm}
%\centerline{\tenrm and}
%\vspace{0.3cm}
%\centerline{\tenrm SECOND AUTHOR'S NAME}
%\baselineskip=13pt
%\centerline{\tenit Group, Company, Address, City, State ZIP/Zone, Country}

\vspace{0.9cm}
\abstracts{We examine string (vortex) formation
at a quench for a weakly-coupled global U(1) theory  when the
excitation spectrum is non-relativistic.    It  is so similar to
vortex production in the corresponding relativistic plasma as to
reinforce arguments for the similarity of vortex production in the
early universe and in low-temperature many-body physics.
}

\vfil
%\vspace{0.8cm}
\twelverm   %modified by CLee 23/07/93
\baselineskip=14pt
%\section{General Appearance}
\vspace*{-0.7cm}
%\subsection{Typeset Scripts}
\vspace*{-0.35cm}
%\vglue 0.4cm
\vglue 0.3cm
%\leftline{\twelveit 1.2. Section Headings}
\vglue 0.4cm
\vglue 1pt
%Section headings are to be in upper and lower case letters, and

% *** Start of R.J.R. main text *****************************************

\section{\bf Introduction}

For many years it has been argued by Kibble\cite{kibble1} and
others\cite{shellard}
that the large-scale structure of
the universe can be attributed to cosmic strings formed at phase transitions
in the Grand Unification era.
Unfortunately, given the unlikely event of observing a cosmic string
(vortex) directly it
is difficult to make the case compelling.

However,
the formation of topological defects like vortices during symmetry-breaking
phase
transitions is not unique to the early universe but generic to many physical
systems.
In particular, recent experiments on the
production
of vortices in superfluid $^{4}He$ \cite{lancaster} and $^{3}He$
\cite{helsinki}
have excited considerable interest. A priori, we might
expect the early universe to have little in common with many-body
systems.  Its highly relativistic regime
is characterised by initial temperatures (or
energies) $T\gg m$
for all particle masses $m$ (in units $k_{B}=c=1$), whereas
low temperature non-relativistic media are characterised by $T\ll m$, to
freeze out antiparticles, but with $T\simeq \mu_{{\rm nr}}$, the
non-relativistic chemical potential.
Nonetheless, the proposition  has been made by the experimental groups
concerned that these, and
other experiments\cite{lancaster2}, may provide an insight on the early
universe.  In
this they have been championed by some theoretical astroparticle
physicists, most notably Zurek\cite{zurek1},
one of whose most recent  articles\cite{zurek2} was, indeed, titled
{\it Cosmological Experiments in Superfluids and Superconductors}.

In this talk I shall indicate how comparisons between
vortex production in these very different
environments can be more than superficial analogies by sketching out
some preliminary quantitative steps.
The talk has three components.

Firstly, I shall remind you of the generalities of vortex production
and of the way a network of simple $Z$-vortices is characterised by
field configuration probabilities.
The next step is to show how, by varying the chemical potential, it
is possible to interpolate between a relativistic scalar
theory (which, in curved space-time, could be of the early universe), and a
non-relativistic field theory of
condensed bosons.
Finally, these ideas will be given a concrete realisation in a model
of $Z$-vortex formation by unstable long-wavelength  Gaussian
fluctuations.

The reader seeking greater detail will find it in recent work by
myself and my collaborators,
Tim Evans and Alasdair Gill\cite{alray,timray,altimray,altimray2},
from which this talk is drawn.

\section{\bf Vortex Production and Vortex Distributions}

What encourages us in the hope that vortex production in the early
universe and the laboratory has close parallels is that
the most plausible production mechanism refers to
neither.
We recapitulate it now for the simplest theory permitting vortices,
that of a complex scalar field $\phi ({\bf x},t)$.  The complex order
parameter of the theory is $\langle\phi\rangle = \eta e^{i\alpha}$ and the
theory possesses
a global $O(2)$ symmetry that we take to be spontaneously broken
at a phase transition.  The field $\phi$ could
be either a complex non-relativistic order field appropriate to
a superfluid or a relativistic field in the early universe.

Initially, we take the system to be in the symmetry-unbroken
(disordered) phase.  We have no
reason to choose any particular initial field configuration, beyond
the requirement that the field is distributed about $\phi = 0$
with zero mean.
The simplest assumption is that, beginning at some time $t = t_0$,
the $O(2)$ symmetry of the ground-state (vacuum) is broken by a rapid change in
the
environment inducing an explicit time-dependence in the field
parameters. Once this quench is completed the $\phi$-field potential
$V(\phi ) = -a|\phi |^{2} + b|\phi |^{4}$ is
taken to have the symmetry-broken form $a > 0,\,\, b > 0$
of the familiar `wine-bottle' bottom.
The ground-state manifold
(the circle $S^{1}$, labelled by the phase $\alpha$ of $\langle\phi\rangle$) is
infinitely connected and the theory possesses {\it global} strings or
vortices, labelled by a winding number $n\in Z$.

These strings cost considerable energy to produce.  That they should appear
at all follows from a general argument,  due
to Kibble\cite{kibble1}, which goes as follows. During the
transition, the complex scalar field begins to
fall from the false ground-state into the true ground-state,
choosing a point on the ground-state manifold at each
point in space.
For continuous
transitions, for which this collapse to the true ground-state
occurs by phase separation,
the resulting field configuration is expected to be one of
domains within each of which the scalar field has relaxed to a constant
ground-state value.

If this is so, then the
requirements of continuity and single valuedness will sometimes force the field
to
remain in the false ground-state between some of the domains. For example, the
phase of
the field may change by an integer multiple of $2 \pi$ on going round
a loop in space. This requires at least one
{\it zero} of the field within the loop, each of which has topological
stability
and characterises a vortex passing through the
loop.   The density of strings is then closely linked to the number
of effective domains and the evolution of this density is,
correspondingly, linked to the nature of the domain growth.
When the phase transition is complete and there is no longer sufficient
thermal energy available for the field to fluctuate into the false
ground-state, the topological defects are frozen into the
field.

Consider an ensemble of systems evolving from one of a set of disordered
states whose relative probabilities are known,
to an ordered state as indicated above.
If the phase change begins
at time $t_{0}$ then, for $t > t_{0}$, it is  possible in principle
to calculate the probability $p_{t}[\Phi]$ that the complex field
$\phi ({\bf x}, t)$ takes the value $\Phi ({\bf x})$ at time $t$.
Throughout, it will be convenient to decompose $\Phi$ into
real and imaginary parts as $\Phi  = \frac{1}{\sqrt{2}}(\Phi_{1} +
i\Phi_{2})$ (and $\phi$ accordingly).  This is because we wish to
track the field as it falls from the unstable ground-state hump at the centre
of the potential to the ground-state manifold in
Cartesian field space.

The calculation of $p_{t}[\Phi]$ will be performed later in our simple model.
For the
moment, consider it given.  The question is, how can we infer the
string densities and the density correlations from  $p_{t}[\Phi]$?
That we can calculate them at all is a consequence of the fact,
noted earlier, that the
string core is a line of field zeroes.
This is equally true for both relativistic and non-relativistic
$O(2)$ theories.
The zeroes of $\Phi_{a}$ ($a$=1,2) which define the vortex positions
form either closed loops or `infinite' string  i.e.string that does
not intersect itself.
Following Halperin\cite{halperin} we define the {\it
topological line density} ${\vec{\rho}}(\bf r)$ by
\ben
{\vec{\rho}}({\bf r}) = \sum_{n}\int ds \frac{d{\bf R}_{n}}{ds}
\delta^{3} [{\bf r} - {\bf R}_{n}(s)].
\een
In (2.1) $ds$ is the incremental length along the line of zeroes ${\bf
R}_{n}(s)$ ($n$=1,2,.. .) and $\frac{d{\bf R}_{n}}{ds}$ is a unit
vector pointing in the direction which corresponds to positive
winding number.  Only winding numbers $n = \pm 1$ are considered,
with higher winding numbers understood as describing multiple zeroes.
If $dA_j$ is an incremental two-dimensional surface containing the
point ${\bf r}$, whose normal is in the $j$th direction, then
$\rho_{j}({\bf r})$ is the net density of strings (i.e the density of
strings {\it minus} the density of antistrings on $dA_j$).

Ensemble averaging $\langle F[\Phi ]\rangle_{t}$ at time $t$ is understood
as averaging over the field probabilities
$p_{t}[\Phi ]$ as
\ben
\langle F[\Phi ]\rangle_{t} = \int {\cal D}\Phi\; p_{t}[\Phi ]\;
F[\Phi ].
\een
In general this ensemble averaging is not
thermal averaging since, out of equilibrium, we have no Boltzmann
distribution.
We shall only consider situations in which
\ben
\langle\rho_{j}({\bf r})\rangle_{t} = 0,
\een
i.e. an equal likelihood of a string or an antistring passing
through an infinitesimal area.  However, the line density
correlation functions
\ben
C_{ij}({\bf r} ;t) = \langle\rho_{i}({\bf r})\rho_{j}({\bf 0})\rangle_{t}
\een
will be non-zero, and give information on the persistence length of
strings.

It follows that, in terms of the zeroes of $\Phi ({\bf r})$,
$\rho_{i}({\bf r})$ can be written as
\ben
\rho_{i}({\bf r}) = \delta^{2}[\Phi ({\bf r})]\epsilon_{ijk}\partial_{j}
\Phi_{1}({\bf r}) \partial_{k}\Phi_{2}({\bf r}),
\label{rho}
\een
where
$\delta^{2}[\Phi ({\bf r})] = \delta[\Phi_{1} ({\bf r})] \delta[\Phi_{2}
({\bf r})]$.
The coefficient of the $\delta$-function in Eq.\ref{rho} is
the Jacobian of the transformation from line zeroes to field zeroes.
 It permits us to define a further line density that we shall also
find useful, the {\it total line density} $\bar{\rho}({\bf r})$
\ben
\bar{\rho_{i}}({\bf r}) = \delta^{2}[\Phi ({\bf r})]|\epsilon_{ijk}\partial_{j}
\Phi_{1}({\bf r}) \partial_{k}\Phi_{2}({\bf r})|.
\label{rhobar}
\een
Unlike the case for $\rho_{i}({\bf r})$
\ben
n(t) = \; \langle\bar{\rho_{i}}({\bf r})\rangle_{t} \; > 0
\een
and measures the {\it total} string density in the direction $i$, without
regard to string orientation.  The isotropy of the initial state
guarantees that $n(t)$ is independent of the direction $i$.
We note that the Jacobian factor multiplying the field
$\delta$-functions in Eq.\ref{rho} and Eq.\ref{rhobar} guarantees that
random field zeroes with no vorticity will not be counted.

In general, the best that we can do is write $\langle\bar{\rho_{i}}({\bf
r})\rangle_{t}$ as
\bea
\langle\bar{\rho_{i}}({\bf
r})\rangle_{t} &=& \int {\cal D}\Phi\; p_{t}[\Phi ]\;
\delta^{2}[\Phi ({\bf r})]|\epsilon_{ijk}\partial_{j}
\Phi_{1}({\bf r}) \partial_{k}\Phi_{2}({\bf r})|
\nonumber
\\
&=& \int {\cal D}\alpha\,\,
\langle |\epsilon_{ijk}\partial_{j}
\Phi_{1}({\bf r}) \partial_{k}\Phi_{2}({\bf r})|\,
e^{i\int d{\bf x}\,\alpha_{a}\Phi_{a}}\rangle_{t}.
\label{robart}
\eea

The simple model that we shall propose later assumes that $p_{t}[\Phi ({\bf
r})]$ is Gaussian.
In this  case the system is solvable, as we shall now show.
Let us assume that
\bea
\langle\Phi_{a}({\bf r})\rangle_{t} &=& 0,
\nonumber
\\
\langle\Phi_{a}({\bf r})\partial_{j}\Phi_{b}({\bf r})\rangle_{t} &=& 0,
\label{g1}
\eea
and, further, that
\bea
\langle\Phi_{a}({\bf r})\Phi_{b}({\bf r}')\rangle_{t} &=& W_{ab}(|{\bf r} -{\bf
r} '|;t)
\nonumber
\\
&=& \delta_{ab} W(|{\bf r} -{\bf r} '|;t),
\label{g2}
\eea
is diagonal.  All other connected correlation functions are taken to
be zero.
Then all ensemble averages are given in terms of $W(r;t)$, where $r
= |{\bf r}|$.  In particular, $\langle\bar{\rho_{i}}({\bf
r})\rangle_{t}$ separates as
\ben
\langle\bar{\rho_{i}}({\bf r})\rangle_{t} =
\langle \delta^{2}
[\Phi ({\bf r})]\rangle_{t}\;\langle |\epsilon_{ijk}\partial_{j}
\Phi_{1}({\bf r}) \partial_{k}\Phi_{2}({\bf r})|\rangle_{t}
\een
or, equivalently, from Eq.\ref{robart}
\ben
\langle\bar{\rho_{i}}({\bf
r})\rangle_{t} = \int {\cal D}\alpha\,\,
\langle |\epsilon_{ijk}\partial_{j}
\Phi_{1}({\bf r}) \partial_{k}\Phi_{2}({\bf r})|\rangle_{t}\,\langle
e^{i\int d{\bf x}\,\alpha_{a}\Phi_{a}}\rangle_{t}.
\een
It follows\cite{halperin}, on first performing the $\alpha$ integration
in the second factor, that
\ben
n(t) =
\frac{1}{2\pi}\biggl |\frac{W''(0;t)}{W(0;t)}\biggr |,
\label{ni}
\een
where primes on $W$ denote differentiation with respect to r.
Thus
\ben
n(t) = O\biggl (\frac{1}{\xi^{2}}\bigg ),
\label{n}
\een
where $\xi$ is the length at time $t$ that sets the scale in $W(r;t)$.

The density-density correlation
functions
\ben
C_{ij}({\bf r} ;t) = A(r;t)\delta_{ij} +
B(r;t)\biggl(\frac{r_{i}r_{j}}{r^{2}} - \delta_{ij}\biggr),
\label{ddc}
\een
have a much more complicated realisation\cite{halperin} in terms of  $W$ and
its
derivatives.
We shall be looking for signs of anticorrelation, which enables us
to determine the persistence length of strings in the network (i.e.
how bendy they are).  The bendier, the more string that will occur
in small loops.  This is important, since large scale structure
formation in the universe requires a certain amount of infinite string.

Nothing that we have said so far discriminates between vortices in a
relativistic or a non-relativistic medium.  This distinction appears
in the definition of $\langle ...\rangle_t$, to which we now turn.

\section{\bf The Relationship Between Relativistic and Non-Relativistic
Media}

In order to compare vortex formation in relativistic and
non-relativistic media it is convenient to develop the
formalism of the relativistic field to accomodate both.  For simplicity we
assume in each
case that the system is initially in {\it equilibrium} in the
disordered phase.  The interpolation between relativistic and
non-relativistic regimes is then
effected by introducing the
chemical potential $\mu$ into the relativistic theory, coupled to the conserved
charge $Q$
(particle number minus antiparticle number) arising from the $O(2)$ symmetry.
Provided $\mu$ is small in
comparison to a particle mass the introduction of such
a potential will have little effect on a phase transition for the
relativistic theory initiated by quenching the system from a high
temperature $T_{0}$ to, effectively, zero temperature.
However, on increasing $\mu$ prior to the transition it becomes more costly to
produce
antiparticles and, if the initial temperature $T_{0}$ is decreased to a
value much less than $\mu$, such
antiparticles as are produced are frozen out.  The system is
then one of non-relativistic particles at a temperature much less than the
particle rest
masses. In this
nonrelativistic regime the transition is induced by a quench in
$\mu$ itself or equivalently, since $\mu$ determines the density, by
a density (pressure) quench.

There are two separate problems.  The first is to show explicitly
how the non-relativistic limit occurs.  The second is to relate this
non-relativistic limit (of a relativistic field) to a conventional
a priori non-relativistic field theory of many-body physics.
However, before we can do either we need to discuss non-equilibrium
field dynamics.

\subsection{Field Dynamics with a Chemical Potential}

{}From the viewpoint given above the chemical potential is seen as determining
the initial conditions for the subsequent dynamics, for which
 we adopt the closed time path
method (Schwinger-Keldysh formalism)\cite{schwinger,mahantapa,keldysh,Go4}.  We
have already assumed
that, at the initial time $t_{0}$, we
are
 in a disordered state
with $\langle \phi \rangle = 0$.
Our ignorance is parametrised by the probability
distribution  $p_{t_0}[\Phi]$ that, at time $t_0$, $\phi(t_0,{\bf x}) =
\Phi({\bf x})$.
For the moment we take it as given.  Whether we are in a
relativistic or non-relativistic regime is largely encoded in $p_{t_0}[\Phi]$.
The subsequent, essentially generic,  non-equilibrium field evolution is driven
by a
rapid change
in the environment.  Specifically, for $t > t_{0}$ the action for the field is
taken to be
\ben
S[\phi] = \int d^{4}x \biggl (
\frac{1}{2} \partial_{\mu} \phi_a \partial^{\mu} \phi_a - \frac{1}{2}
m^{2}(t) \phi_a^2 - \frac{1}{4} \lambda (t) (\phi_a^2)^2
\biggr ).
\label{St}
\een
where $m(t)$, $\lambda (t)$ describe the evolution of the parameters
of the theory under external influences, to which the field responds.
There is no penalty in using the Lorentz covariant Eq.\ref{St}, even
for non-relativistic media. If, initially, there are no
antiparticles at low temperature, there will remain no antiparticles.
As with $\Phi$, it is convenient to decompose $\phi$ in terms of two massive
real scalar fields $\phi_a$, $a=1, 2$ as $\phi = (\phi_1 + i\phi_2)/\sqrt{2}$,
in terms of which $S[\phi ]$ shows a global $O(2)$ invariance, broken by the
mass
term if $m^{2}(t)$ is negative.

The change of phase that begins at time $t_{0}$ will, by the
mechanism indicated earlier, lead to the appearence of vortices.
We saw in the previous section that the vortex distributions at later
times
 $t_f>t_0$ can be read
off from the probability $p_{t_f}[\Phi_f]$ that the
measurement of $\phi$ will give the value $\Phi_f$.  The evolution
of $p_{t}[\Phi ]$ from $t_{0}$ to $t_{f}$
is most simply
written as a closed time-path integral
in which the field $\phi$ is integrated along the closed path $C_+ \oplus C_-$
of Fig.1, where $\phi =\phi_+$ on $C_+$ and $\phi= \phi_-$ on $C_-$.
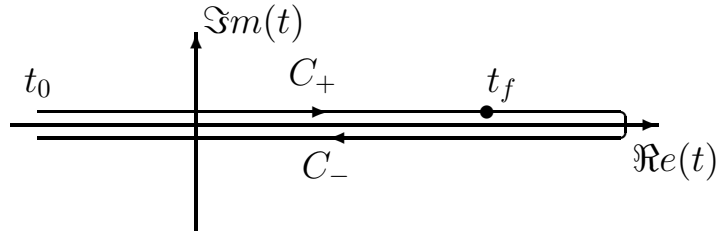
\begin{figure}[htb]
\begin{center}
\setlength{\unitlength}{0.5pt}
\begin{picture}(495,200)(35,600)
\put(260,640){\makebox(0,0)[lb]{\large $C_-$}}
\put(185,750){\makebox(0,0)[lb]{\large $\Im m(t)$}}
\put(250,710){\makebox(0,0)[lb]{\large $C_+$}}
\put(510,645){\makebox(0,0)[lb]{\large $\Re e (t)$}}
\put(400,705){\makebox(0,0)[lb]{\large $t_f$}}
\put( 50,705){\makebox(0,0)[lb]{\large $t_0$}}
\thicklines
\put(400,690){\circle*{10}}
\put( 40,680){\vector( 1, 0){490}}
\put( 60,690){\vector( 1, 0){220}}
\put(280,690){\line( 1, 0){220}}
\put(500,680){\oval(10,20)[r]}
\put(500,670){\vector(-1, 0){220}}
\put(280,670){\line(-1, 0){220}}
%\put( 60,670){\vector( 0,-1){110}}
%\put( 60,560){\line( 0,-1){ 60}}
\put(180,600){\vector( 0, 1){150}}
\end{picture}
\end{center}
\caption{The closed timepath contour $C_+ \oplus C_-$.}
\end{figure}
If ${\cal D} \phi_{\pm} = \prod_{a=1}^2 {\cal D} \phi_{\pm ,a}$ and spatial
labels are suppressed then
\ben
p_{t_f}[\Phi_f] = \int {\cal D} \Phi\, p_{t_0}[\Phi]
\int_{\phi_{\pm} (t_0) = \Phi} {\cal D} \phi_+  {\cal D}
\phi_- \, \delta [ \phi_+(t_f) - \Phi_f ] \, \exp \biggl \{ i \biggl (
S[\phi_+] - S[\phi_-] \biggr ) \biggr \}.
\een
where $\delta [ \phi_+(t) - \Phi_f ]$ is a delta functional, imposing
the constraint $\phi_+({\bf x},t) = \Phi_f ({\bf x})$ for each ${\bf x}$.
This is no more than the statement that, for a given initial state,
 the probability amplitude is
given by the integration along $C_{+}$, and its complex conjugate
(which, when multiplied with it, gives the probability) is given by
the integration back along $C_{-}$.
The $\pm$ two-field notation is misleading in that it suggests that the
$\phi_{+}$ ($=\phi_{a,+}$) and $\phi_{-}$ fields are decoupled.  That this is
not so
follows immediately from the fact that $\phi_{+}(t_f) =\phi_{-}(t_f)$.

To return to the initial conditions, the simple
analytic results of the previous section could only be achieved if $p_{t}[\Phi
]$ were
Gaussian and, therefore, that $p_{t_0}[\Phi]$ itself be
Gaussian.  This may not be unrealistic.
The simplest such distribution has
$\Phi$  Boltzmann distributed at time $t_0$ at a
temperature of $T_0 = \beta_0^{-1}$ and chemical potential $\mu$
according to a {\it free}-field Hamiltonian
$H_0$,
which we take as
\ben
H_{0} = \int d^{3}x \biggl [ \frac{1}{2}\pi_{a}^{2} +
\frac{1}{2} (\nabla\phi_{a} )^{2} + \frac{1}{2} m^2 \phi_{a}^2
\biggr ].
\label{H_{0}}
\een
where $\pi_{a} = \dot{\phi_{a}}$ in real time.
We stress that Gaussian does not necessarily mean {\it free}.
Most simply, particles need to interact before they can
equilibriate.  More importantly, the free-field Gaussian approximation adopted
here
can be extended to include interactions self-consistently
in a Hartree approximation\cite{boyanovsky}.

Since chemical potentials are not usually relevant to
relativistic bosons a few words are in order.
The $O(2)$ invariance of $H_0$ leads to a conserved Noether current with
conserved charge
\bea
Q &=& \int d^{3}x\; (\phi_{2}\pi_{1} - \pi_{2}\phi_{1})
\nonumber
\\
&=& \int d^{3}x\; (\phi_{2}\dot{\phi_{1}} - \dot{\phi_{2}}\phi_{1}).
\label{Q}
\eea
The numerical value of $Q$ is the number of particles minus the
number of antiparticles.
The thermal probability distribution $p_{t_0}[\Phi]$ is thus taken
to be
\ben
p_{t_0}[\Phi] = \langle \Phi,t_0 | e^{- \beta_{0} (H_0 - \mu Q)} | \Phi,t_0
\rangle .
\een
We note that interactions with weak coupling will not change
$p_{t_0}[\Phi]$ significantly.

There are two ways to proceed.  The first, which to us is the
most natural, accepts $H_0$ as
determining the temporal evolution of the fields, and relegates $\mu$
to the initial boundary conditions on the fields.
{}From this viewpoint, $p_{t_0}[\Phi]$ can be written as the
imaginary-time path integral
\ben
p_{t_0}[\Phi] =
\int_{B_{\mu}[\Phi]} {\cal D} \phi
\exp \biggl \{ i S_0 [\phi ] \biggr \},
\label{poo}
\een
for a corresponding action
\ben
S_{0}[\phi] = \int d^{4}x \biggl [
\frac{1}{2} (\partial_{\nu} \phi_{a} )(\partial^{\nu} \phi_{
a} ) - \frac{1}{2} m^2 \phi_{a}^2
\biggr ].
\label{S_{0}}
\een
The
boundary condition $B_{\mu}[\Phi]$ incorporates the chemical potential.
In terms of the $Q$ eigenstates,
the complex field $\phi = (\phi_{1} +
i\phi_{2})/\sqrt{2}$ and its adjoint, $B_{\mu}[\Phi ]$ is
\ben
B_{\mu}[\Phi ]:\,\phi (t_0) = \Phi =
e^{-\beta_{0}\mu q}\phi (t_0-i \beta_0),
\label{bmu}
\een
where $q = 1$ is the $\phi$-field  eigenvalue of $Q$ and the
time-integral in Eq.\ref{S_{0}} is taken in imaginary time from
$t_0$ to $t_0 -i\beta_{0}$.
In terms of
the same complex fields we have
\ben
S_{0}[\phi] = \int d^{3}x [(\partial_{\nu} \phi^{*})(\partial^{\nu}\phi ) - m^2
\phi^{*}\phi ]
\een
and, with $\pi = (\pi_{1} - i\pi_{2})/\sqrt{2}$,
\ben
H_{0} = \int d^{3}x [ \pi^{*}\pi +(\nabla\phi^{*} )(\nabla\phi )
 + m^2 \phi^{*}\phi ].
\label{Hc2}
\een
However, for calculating vortices
a decomposition in terms of $\phi_{a}, \pi_{a}$ is usually preferable.

Although we are just setting an initial condition the effect is, inevitably, to
give an
action $S_{0}[\phi]$ of the form of $S[\phi]$ of Eq.\ref{St}.
This permits the interpretation that the action  $S[\phi]$ is
valid for all times $t$, with the proviso that the system is in thermal
equilibrium
for  $t<t_0$, during
which period the mass $m(t)$ takes the constant value $m$ and $\lambda(t)
= 0$.

On relabelling the integration variable $\phi$ of Eq.\ref{poo} by $\phi_3$,
we now have the explicit form for $p_{t_f}[\Phi_f]$:-
\begin{eqnarray}
\nonumber
p_{t_f}[\Phi_f] &=& \int {\cal D} \Phi \int_{B_{\mu}[\Phi]}
{\cal D} \phi_3 \,  e^{i S_0[\phi_3]}
\int_{\phi_{\pm}(t_0) = \Phi}
 {\cal D} \phi_+ \Phi {\cal D} \phi_- \,
e^{i ( S[\phi_+] - S[\phi_-] ) } \delta [ \phi_+(t_f) - \Phi_f ]
\\
&=& \int_{B_{\mu}} {\cal D} \phi_3 {\cal D} \phi_+ {\cal D} \phi_- \, \exp
\biggl \{
i S_0[\phi_3] + i ( S[\phi_+] - S[\phi_-] )
\biggr \} \,
\delta [ \phi_+(t_f) - \Phi_f ],
\end{eqnarray}
where the boundary condition $B_{\mu}$ is now (in terms of the field
combinations $\phi = (\phi_{1} +
i\phi_{2})/\sqrt{2}$)
\ben
B_{\mu}:\,\,\phi_{+}(t_0) =
e^{-\beta_{0}\mu q}\phi_3(t_0- i \beta_0).
\een
More succinctly, $p_{t_f}[\Phi_f]$ can be written as the time ordering of a
single field doublet:-
\ben
p_{t_f} [ \Phi_f] = \int_{B_{\mu}} {\cal D} \phi \, e^{i S_C [\phi]} \, \delta
[
\phi_+ (t_f) - \Phi_f ],
\label{pf}
\een
along the contour $C=C_+ \oplus C_- \oplus C_3$ of Fig.2, extended to include a
third imaginary leg, where $\phi$ takes the values $\phi_+$, $\phi_-$
and $\phi_3$ on $C_+$, $C_-$ and $C_3$ respectively, for which $S_C$
is $S[\phi_+]$, $S[\phi_-]$ and $S[\phi_3]$, for which last case
$m(t) = m, \lambda (t) = 0$, $t\in C_{3}$.
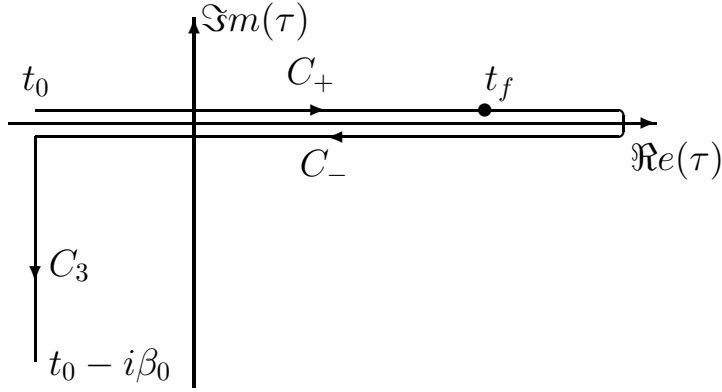
\begin{figure}[htb]
\begin{center}
\setlength{\unitlength}{0.5pt}
\begin{picture}(495,280)(35,480)
\put( 70,565){\makebox(0,0)[lb]{\large $C_3$}}
\put(260,640){\makebox(0,0)[lb]{\large $C_-$}}
\put(185,750){\makebox(0,0)[lb]{\large $\Im m(\tau)$}}
\put(250,710){\makebox(0,0)[lb]{\large $C_+$}}
%\put(285,685){\makebox(0,0)[lb]{\large $0$}}
\put(510,645){\makebox(0,0)[lb]{\large $\Re e (\tau)$}}
\put(400,705){\makebox(0,0)[lb]{\large $t_f$}}
\put( 50,705){\makebox(0,0)[lb]{\large $t_0$}}
\put(70,490){\makebox(0,0)[lb]{\large $t_0-i \beta_0$}}
\thicklines
\put(400,690){\circle*{10}}
\put( 40,680){\vector( 1, 0){490}}
\put( 60,690){\vector( 1, 0){220}}
\put(280,690){\line( 1, 0){220}}
\put(500,680){\oval(10,20)[r]}
\put(500,670){\vector(-1, 0){220}}
\put(280,670){\line(-1, 0){220}}
\put( 60,670){\vector( 0,-1){110}}
\put( 60,560){\line( 0,-1){ 60}}
\put(180,480){\vector( 0, 1){280}}
\end{picture}
\end{center}
\caption{A third imaginary leg}
\end{figure}
%Henceforth we drop the suffix $f$ on $\Phi_f$ and take the
%origin in time from which the evolution begins as $t_0 =0$.

As a final step in these formal manipulations we see that expression
Eq.\ref{pf} enables us to write the $\Phi$-field ensemble averages
$\langle ...\rangle_{t}$ in terms of the $\phi$-field Wightman
functions.
Consider the generating
functional:-
\ben
Z_{\mu}[j_+,j_-,j_3] = \int_{B_{\mu}} {\cal D} \phi \, \exp \biggl \{ i
S_C[\phi] +
i \int j \phi \biggr \},
\label{Z}
\een
where $\int j \phi$ is shorthand for:-
\ben
\int j \phi \equiv \int_0^{\infty} dt \, \,  [ \, j_+(t) \phi_+(t) - j_-
\phi_-(t) \, ] \, + \int_0^{-i \beta} j_3(t) \phi_3(t) \, dt,
\een
omitting spatial arguments. Then introducing $\alpha_a({\bf x})$ where
$a=1, 2$, we find:-
\bea
p_{t_f} [\Phi] &=& \int {\cal D} \alpha \int_{B_{\mu}} {\cal D } \phi \,\,
\exp \biggl \{i
S_C[\phi] \biggr \} \,\,
\exp \biggl \{ i\int d^{3} x \alpha_a({\bf x}) [ \phi_+(t_f,{\bf
x}) - \Phi({\bf x}) ]_a \biggr \}
\nonumber
\\
&=& \int {\cal D} \alpha \, \, \exp \biggl \{ -i \int \alpha_a \Phi_a
\biggr \} \,Z_{\mu}[\overline{\alpha},0,0],
\eea
where $\overline{\alpha}$ is the source $\overline{\alpha}({\bf x},t) =
\alpha({\bf x}) \delta (t-t_f)$. As with ${\cal D} \phi$, ${\cal D}
\alpha$ denotes $\prod_1^N {\cal D} \alpha_a$.

Ensemble averages are now expressible in terms of $Z_\mu$.  Of
particular relevance,
\bea
W_{ab}(|{\bf r} -{\bf r} '|;t) &=&
\langle\Phi_{a}({\bf r})\Phi_{b}({\bf
r}')\rangle_{t}
\nonumber
\\
&=& \langle\phi_{a}({\bf r},t)\phi_{b}({\bf
r}',t)\rangle,
\eea
the equal-time Wightman function with the given thermal boundary conditions.
Because of the time evolution
there is no time translation invariance in the double time label.

Not surprisingly, $p_{t} [\Phi ]$ can only be calculated explicitly in
very simple circumstances, like our Gaussian approximation, but
before we do that we still have to extract the relativistic limit.
Since the chemical potential is embedded in the equilibrium boundary
conditions and, like temperature, can only be defined in
equilibrium, this is essentially an equilibrium problem and, for
the moment, we forget the dynamics.

\subsection{Deriving the Non-Relativistic Limit}

The expression Eq.\ref{pf} is valid for all $\mu$ and all $T_{0}$
but, as it stands, is not
sympathetic to the isolation of a non-relativistic limit in a form
with which we are familiar.
Nonetheless, the extraction of a non-relativistic regime from it is
not difficult.
Instead of working with $p_{t_{0}}[\Phi ]$ directly, we can work
with the partition function $Z_{\mu}$, as we saw in the previous section.
The partition function for this equilibrium theory is (for doublet
sources $j_{a}$ restricted to the $C_{3}$-contour)
$Z_{\mu}[0,0,j_{3}]$ of Eq.\ref{Z}, written
\ben
Z_{\mu}[j]
\equiv Z_{\mu}[0,0,j]
= \int_{B_{\mu}} {\cal D} \phi \, \exp \biggl \{ i S_{0}[\phi] +
i \int j_{a} \phi_{a} \biggr \},
\label{Z_{0}}
\een
where $S_{0}[\phi]$ is given in Eq.\ref{S_{0}} and we integrate only
along the contour $C_{3}$.
%$B_{\mu}$ now means $\phi (t_0) =
%e^{-\beta_{0}\mu q}\phi (t_0- i \beta_0)$.
  As before, ${\cal D}\phi\equiv{\cal
D}\phi_{1}{\cal D}\phi_{2}$.  On rotating to Euclidean time $\tau =
it$, $Z_{\mu}[j]$ becomes
 \ben
Z_{\mu}[j] = \langle\exp \biggl \{\int j_{a} \phi_{a} \biggr \}\rangle =
 \int_{B_{\mu}} {\cal D} \phi \, \exp \biggl \{ - S_{0,E}[\phi] -
\int j_{a} \phi_{a} \biggr \},
\label{Z_{E}}
\een
where $S_{0,E}[\phi]$ is the relativistic Euclidean action
\ben
S_{0,E}[\phi ] = \int_{0}^{\beta_{0}} d\tau \int d^{3}x \biggl [
\frac{1}{2}\dot{\phi}_{a}^{2} +
\frac{1}{2} (\nabla\phi_{a} )^{2} + \frac{1}{2} m^2 \phi_{a}^2
\biggr ].
\label{S_{0,E}}
\een
and the sum is taken over fields $\phi = (\phi_1 +
i\phi_2)/\sqrt{2}$
 satisfying the boundary condition $B_{\mu}$: $\phi ({\bf
x}, \tau) = e^{-\beta_{0}\mu}\phi ({\bf
x}, \tau-\beta_{0})$ in imaginary
time.  The dot now means differentiation with
respect to $\tau$, $\dot{\phi_{a}}= \partial_{\tau}\phi_{a}$.

The bracket $\langle ...\rangle$ here denotes the thermal average
\ben
\langle F[\phi]\rangle = Tr\{e^{-\beta_{0} (H_{0} - \mu Q)}\; F[\phi]\}.
\label{chemav}
\een
where $H_{0}$ is given in Eq.\ref{H_{0}}.
Since it is time-independent we have dropped the $t$-suffix.

$Z_{\mu}[j]$ can be calculated as it stands. However, because
of the chemical potential in the boundary conditions it is
not immediately obvious how to identify the phase of the system.
This is
clarified by adopting a second approach to chemical potentials, in
which we transfer the chemical potential term $\mu
Q$ to the
Hamiltonian, to create an effective Hamiltonian $H_0 - \mu Q$.
Although $H_0 - \mu Q$ no longer generates time translations of the
physical fields $\phi$ and $\phi^{*}$, it does generate time
translations of the effective field $\tilde{\phi}(t) = e^{i\mu t}\phi
(t)$ (and its adjoint).  To
see the advantages of removing $\mu$ from the boundary conditions we express
$Z_{\mu}$ of
Eq.\ref{Z_{0}} not as a sum over $\phi$-field
histories but as a
sum over $\tilde{\phi}$-field histories.
Although not usually posed in this way the details are
well-understood.  For
example, see Kapusta\cite{kapusta}, Haber $\&$ Weldon\cite{art},  and Bernstein
et al.\cite{dodelson}.

With the Jacobian unity,
$Z_{\mu}[j]$ becomes
 \ben
Z_{\mu}[j] =
 \int_{B_{0}} {\cal D} \tilde{\phi} \, \exp \biggl \{ - S_{\mu}[\tilde{\phi}] -
\int j_{a} \tilde{\phi_{a}} \biggr \},
\label{Zmu}
\een
where $S_{\mu}[\phi ]$ is now the effective action
\ben
S_{\mu}[\phi ] = \int_{0}^{\beta_{0}} d\tau \int d^{3}x \biggl [
\frac{1}{2}\dot{\phi}_{a}^{2} +
\frac{1}{2} (\nabla\phi_{a} )^{2} + \frac{1}{2}m_{0}^2 \phi_{a}^2
+ i\mu (\phi_{2}\dot{\phi_{1}} - \dot{\phi_{2}}\phi_{1})
\biggr ].
\label{S0}
\een
in which
\ben
m_{0}^{2} = m^2-\mu^{2}
\een
It follows from Eq.\ref{bmu} that, with $\mu Q$ bundled with $H_{0}$
the $\tilde{\phi_{a}}$ integrations are now taken
over {\it periodic} configurations with boundary conditions
$B_{0}:\tilde{\phi_{a}}({\bf
x}, \tau) = \tilde{\phi_{a}}({\bf
x}, \tau-\beta_{0})$ in imaginary
time, with no $\mu$-dependence.
Most importantly, the zeroes of $\tilde{\phi}$ are the zeroes of
$\phi$, leading to identical line densities.

This displacement of $\mu$ from the boundary conditions to the
action has enabled us to replace the classical potential by an
effective potential in which the state of the system is more transparent.
Semiclassically, when $m_{0}^{2} > 0$ (i.e. $\mu^{2} < m^{2}$) the $O(2)$
symmetry is unbroken.
This describes our disordered initial state.
However, once $m_{0}^{2} < 0$ (i.e. $\mu^{2} > m^{2}$) the free theory is
unstable.
%At this level $\mu_{c} = m$ in Fig.1.
This is a signal that the $O(2)$ symmetry is broken and a transition
 to an ordered phase has occurred.  Thus, effectively, it is
$m_{0}^{2}$, rather than $m^2$ that carries the time dependence,
changing from positive to negative at $t = t_{0}$.

Semiclassically, the non-relativistic limit is trivial.  Although we achieve it
by
following the phase boundary $m_{0}^{2} = 0$ from high $T$ and low
$\mu$ to low $T$ and high $\mu$ the decomposition of $m_{0}^{2}$ is
immaterial.  Only in the cross-terms $i\mu (\phi_{2}\dot{\phi_{1}} -
\dot{\phi_{2}}\phi_{1})$ will the change be noticed.
Even this is misleading since, as we shall see, this term plays no
role  in our Gaussian approximation.

All that remains is a problem of interpretation.
Let us first consider the relativistic regime in which the initial
environment is very hot, with $T\gg m\gg
\mu$.
In this case, with $\mu$ irrelevant, the symmetry is broken by a change in
$m^{2}$.
Suppose that, on the completion of the transition, $m_{0}^{2}\simeq
m^{2}$ takes the value $m_{0}^{2} = -M^{2} < 0$.
If, in this relativistic case, the final temperature is very low, there are
then no thermal effects once interactions have been taken into account
and $M$ is a physical parameter, determining the Higgs mass, $m_{H}
= \sqrt{2} M$.
However, because of thermal radiative effects $m_{0}(t\leq 0) = m_{0}$
is not a physical parameter. In practice, there
is no loss in taking it as the effective
scalar field mass
at temperature $T_{0} = \beta_{0}^{-1}$, despite the fact that such
a mass is defined in terms of large scale fluctuation averages (the
effective potential). That is, in the mean-field approximation,
\ben
m_{0}^{2} = -M^{2}\bigg (1-\frac{T^2}{T_{0}^{2}}\bigg ).
\label{T/Tc}
\een
The results will turn
out to be largely independent of $m_{0}$ provided it is comparable to
$M$, which is the case if we do not quench from too close to the transition.
On the other hand, for the non-relativistic theory, for which
$T/m\ll 1$, the parameter $m$ can always be identified with the
boson mass.  In this case we
change the sign of $m_{0}^{2}$ in the
action by a change in $\mu$ from
$\mu^{2} < m^{2}$ to $\mu^{2} > m^{2}$, a density quench.
To preserve uniformity of notation, we take $m_{0}^{2}(t) =
m_{0}^{2}$ when $t\leq t_{0}$ and $m_{0}^{2}(t) = -M^{2} < 0$ once
the transition is complete.
\newpage
\subsection{Extracting the Non-Relativistic Field Theory}

At this stage we could proceed directly to our Gaussian fluctuation
model.  However, it is advantageous to show how the unfamiliar equation
Eq.\ref{S0} can be cast in a form standard to non-relativistic
many-body theory.  In fact,
one of the difficulties of making straightforward comparisons
between early universe physics and many-body physics is this
difference in formalism, a reflection of the freezing out of
antiparticles in the non-relativistic regime.  Thus, for example, a
non-relativistic order field, with its interpretation as a probability
amplitude
for finding particles, is necessarily complex.  As is apparent, our main tools
in relativistic particle physics are path integrals which,
conventionally, are not separated into particle and antiparticle
sectors.  This split has to be made if we are to recover the
familiar non-relativistic as a field theory in its own right.

To do this we rewrite $Z_{\mu}$ of Eq.\ref{Zmu} as a sum over
histories   in the full phase-space of the
fields $\tilde{\phi_{a}}$ {\it and} their conjugates $\tilde{\pi_{a}}$.
As we shall see, it is just this extended sum over
histories that is needed for isolating the non-relativistic limit of
the theory.

As a phase-space integral, $Z_{\mu}[j]$ takes the form
\ben
Z_{\mu}[j] = \int_{B_{0}} {\cal D}\tilde{\pi_{1}}{\cal D}\tilde{\pi_{2}}{\cal
D}\tilde{\phi_{1}}{\cal D}\tilde{\phi_{2}}
\exp \biggl \{ - \tilde{S}_{\mu ,E}[\tilde{\pi}, \tilde{\phi} ] -
\int j_{a} \tilde{\phi_{a}} \biggr \},
\label{Zom}
\een
where
\ben
- -\tilde{S}_{\mu ,E}[\pi, \phi ] = \int_{0}^{\beta_{0}}d\tau\biggl \{
\biggl( \int d^{3}x\; i\pi_{a}\dot{\phi_{a}}\biggr) - \biggl( H_{0}
- - \mu Q \biggr)  \biggr \}.
\een
We have made the integration over the {\it four} field degrees of
freedom explicit.  Although the $\tilde{\phi}-{a}$-fields are
constrained to be periodic, the $\tilde{\pi_{a}}$ integrations are
unconstrained, and
$\tilde{S}_{\mu ,E}$ cannot be identified directly with the `effective'
action in the presence of a chemical potential.  However, the $\tilde{\pi}$
integrations are readily performed to recreate $Z_{\mu}$ of Eq.
\ref{Zmu}.

It has been shown in the work of Kapusta, Bernstein et al.\cite{dodelson} how,
in the limit $m\gg
T_{0}\simeq -\mu_{{\rm nr}}$, where $\mu_{{\rm nr}} = \mu - m$, we recover
the energy spectrum (and free energy) of a non-relativistic Bose gas.
We shall now show that this same phase-space analysis permits us
to recover
 the familiar non-relativistic
partition function, for free bosons,
\ben
Z_{\mu_{{\rm nr}}}[\eta^{*},\eta ] = \int_{b_{0}} {\cal D}\psi^{*} {\cal
D}\psi\;\exp \biggl \{ - S_{\mu_{\rm nr}}[\psi^{*},\psi ]  +
\int (\eta \psi^{*} +\eta^{*} \psi )   \biggr \},
\label{Znr}
\een
where the boundary condition $b_{0}$ denotes simple periodicity of
the fields $\psi ,\psi^{*}$ in imaginary time with period $\beta_{0}$.
In Eq.\ref{Znr} $S_{\mu_{\rm nr}}[\psi^{*},\psi ]$ is the non-relativistic
action
\bea
 S_{\mu_{\rm nr}}[\psi^{*},\psi ] &=& \int_{0}^{\beta_{0}}d\tau\int d^{3}x
\biggl[ \frac{1}{2}(\psi^{*}\dot{\psi} - \dot{\psi^{*}}\psi ) +
\frac{1}{2m}(\nabla\psi^{*})(\nabla\psi) -
\mu_{{\rm nr}}\psi^{*}\psi \biggr]
\nonumber
\\
&=&  \int_{0}^{\beta_{0}}d\tau\int
d^{3}x\,\psi^{*}\biggl[\frac{\partial}{\partial\tau} -
\frac{1}{2m}\nabla^{2} - \mu_{{\rm nr}}
\biggr]\psi
\label{Snr}
\eea
in the order field $\psi$ and its conjugate.

The non-relativistic
field $\psi ({\bf x},t)$ is understood as annihilating a particle
at the point ${\bf x}$, whereas $\psi^{*}({\bf x},t)$ creates a
particle at ${\bf x}$.  Although the particles are distinct from the
antiparticles there are no antiparticles in $Z_{\mu_{{\rm nr}}}$.
$|\psi|^{2}$ measures the local particle density, a role
not available to $|\phi |^{2}$ of the relativistic field theory.  Of course, we
can decompose $\psi$ as $\psi = \frac{1}{\sqrt{2}}(\psi_{1}+
i\psi_{2})$ but, for once, we decline to do so since $\psi_{1}$ and
$\psi_{2}$ have no obvious interpretation.

$Z_{\mu}[j]$ of Eq.\ref{Zmu} and
$Z_{\mu_{{\rm nr}}}[\eta^{*},\eta ]$ of
Eq.\ref{Znr} cannot be identical, coming from different origins, but
they are closely related in the regime $m\gg T_{0}\simeq -\mu_{{\rm nr}}$.
To see this, we rewrite $Z_{\mu}[j]$ of Eq.\ref{Zom} in
terms of the complex fields $\phi ,\phi^{*} ,\pi ,\pi^{*}$ as
\ben
Z_{\mu}[j, j^{*}] = \int {\cal D}\pi^{*}{\cal D}\pi{\cal D}\phi^{*}{\cal
D}\phi
\exp \biggl \{ - \tilde{S}_{\mu ,E}[\pi, \phi ] -
\int (j^{*}\phi + j\phi^{*}) \biggr \},
\label{Zc}
\een
where we have dropped all {\it tilde} labels from the fields and
\ben
- -\tilde{S}_{\mu ,E}[\pi, \phi ] = \int_{0}^{\beta_{0}}d\tau\biggl \{
\int d^{3}x\; \biggl( i\pi\dot{\phi} +  i\pi^{*}\dot{\phi^{*}}  \biggr) -
\biggl( H_{0}
- - \mu Q \biggr)  \biggr \},
\label{Sc}
\een
in which $H_{0}$ is given in Eq.\ref{Hc2} and $Q$ is
now, equivalently
\ben
Q = i\int d^{3}x\; (\phi^{*}\pi^{*} - \pi\phi ).
\een

We could proceed as before, and perform the $\pi ,\pi^{*}$
integrations, which would give us $Z_{\mu}[j]$ of Eq.\ref{Zmu} in
complex field form, but this is nothing new.  Instead, we perform
canonical transformations
from the phase-space fields $\phi ,\phi^{*},\pi ,\pi^{*}$ to a new
set of fields
$\Psi ,\Psi^{*}, \bar{\Psi},\bar{\Psi}^{*}$ that, in the
non-relativistic limit, annihilate and create particles and
antiparticles.
Specifically,
\bea
\Psi &=& \sqrt{\frac{m}{2}}\phi + \frac{i}{\sqrt{2m}}\pi^{*},
\nonumber
\\
\Psi^{*} &=& \sqrt{\frac{m}{2}}\phi^{*} - \frac{i}{\sqrt{2m}}\pi ,
\label{can}
\eea
annihilate and create non-relativistic particles in this limit, whereas
\bea
\bar{\Psi} &=& \sqrt{\frac{m}{2}}\phi^{*} + \frac{i}{\sqrt{2m}}\pi ,
\nonumber
\\
\bar{\Psi}^{*} &=& \sqrt{\frac{m}{2}}\phi - \frac{i}{\sqrt{2m}}\pi^{*},
\label{can2}
\eea
annihilate and create non-relativistic {\it antiparticles}
respectively.

To confirm that this is the case we invert the above.  The field
measure
\ben
{\cal D}\pi^{*}{\cal D}\pi {\cal D}\phi^{*}{\cal D}\phi ={\cal D}\Psi^{*}{\cal
D}\Psi{\cal D}\bar{\Psi}^{*}{\cal
D}\bar{\Psi}
\een
is unchanged.  Inserting these decompositions in $Z_{\mu}[j, j^{*}]$
of Eq.\ref{Zc} (where, for simplicity, we switch off the sources $j,
j^{*}$) gives $Z_{\mu}$ of the form
\ben
Z_{\mu}\equiv Z_{\mu}[0,0] = \int_{B_{0}} {\cal D}\Psi^{*}{\cal D}\Psi{\cal
D}\bar{\Psi}^{*}{\cal
D}\bar{\Psi}\exp\{-\tilde{S}_{\mu ,E}\}
\label{Zpp}
\een
where $\tilde{S}_{\mu ,E}$ of Eq.\ref{Sc} is now a functional of $\Psi
,\bar{\Psi}$, etc., but $B_{0}$ is a boundary condition on the
$\phi$ fields, and {\it not} the $\Psi$ fields.

Then,
on direct substitution $\tilde{S}_{\mu ,E}$ is
\bea
\tilde{S}_{\mu ,E} &=& \int_{0}^{\beta_{0}}d\tau\int
d^{3}x\,\Psi^{*}\biggl[\frac{\partial}{\partial\tau} -
\frac{1}{2m}\nabla^{2} + (m - \mu )\biggr]\Psi
\nonumber
\\
&+& \int_{0}^{\beta_{0}}d\tau\int
d^{3}x\,\bar{\Psi}^{*}\biggl[\frac{\partial}{\partial\tau} -
\frac{1}{2m}\nabla^{2} + (m + \mu )\biggr]\bar{\Psi}
\nonumber
\\
&+&  \int_{0}^{\beta_{0}}d\tau\int d^{3}x\,\frac{1}{2m}
\biggl\{ (\nabla\bar{\Psi})(\nabla\Psi) +
(\nabla\bar{\Psi}^{*})(\nabla\Psi^{*})\biggr\},
\label{Sl}
\eea
after removing total derivatives.
This can be written in terms of the non-relativistic action
$S_{\mu_{\rm nr}}[\psi^{*},\psi ]$ of
Eq.\ref{Snr} as
\ben
\tilde{S}_{\mu ,E} = S_{\mu_{\rm nr}}[\Psi^{*},\Psi ] +
S_{-\mu -m}[\bar{\Psi}^{*},\bar{\Psi} ]
+ S_{0, I}[\Psi^{*},\Psi,\bar{\Psi}^{*},\bar{\Psi}],
\een
where $S_{0, I}$, coupling particles to antiparticles, is the third
term of equation Eq.\ref{Sl} and $\mu_{nr} = \mu -m$.
$S_{\mu_{\rm nr}}[\Psi^{*},\Psi ]$ and
$S_{-\mu -m}[\bar{\Psi}^{*},\bar{\Psi} ]$ each possess an
$O(2)$ invariance, for the individual conservation of particle and
antiparticle number.  The $O(2)$ symmetry of the antiparticle action $S_{-\mu
-m}[\bar{\Psi}^{*},\bar{\Psi}]$
remains unbroken.  It is the $O(2)$ symmetry of the particle
action $S_{\mu_{\rm nr}}$ of
Eq.\ref{Snr}  which leads to non-relativistic vortices on its
breaking.  Away from the non-relativistic
limit  the  $O(2)\otimes O(2)$ symmetry
of $\tilde{S}_{\mu ,E}$ is reduced to the $O(2)$ of the relativistic theory
(the
conservation of particle minus antiparticle number) by $S_{0, I}$.

To see this pattern we observe that, on partial integration, $S_{0, I}$
can be thought of as providing sources for $\bar{\Psi}$ and $\bar{\Psi}^{*}$ in
$Z_{\mu}$ of Eq.\ref{Zpp}, which can be integrated out.
This gives $Z_{\mu}$ as an integral over the non-relativistic
partition with chemical potential $-m-\mu$,
\ben
Z_{\mu}= \int_{B_{0}} {\cal D}\Psi^{*}{\cal D}\Psi
\exp\{-S_{\mu_{\rm nr}}[\Psi^{*},\Psi ]\}\,
Z_{-m-\mu}[-\nabla^{2}\Psi/2m,
- -\nabla^{2}\Psi^{*}/2m]
\label{Zp}
\een
in the terminology of Eq.\ref{Znr} where, again, the periodic boundary
condition is on the $\psi$ fields.  For the free-field case in hand,
$Z_{-m-\mu}$ in the integrand is Gaussian in its sources.
 For the non-relativistic regime, in which
$|\mu - m|\ll m$, whence $-\mu -m\approx -2m$,
the factor $Z_{-m-\mu}\approx Z_{-2m}$ in the integrand of Eq.\ref{Zp} can be
expanded in
powers of momenta.  For particle kinetic energies ${\bf
k}^{2}/2m\ll m$, $Z_{-m-\mu}\approx 1$.
For typical $k = |{\bf k}|$, with $k^{2}/2m = O(T)$, the errors
are $O(k^{2}/m^{2}) = O(T/m)$.
{}From Eq.\ref{can} and Eq.\ref{can2}, we identify
the semi-classical $\phi$ with $\Psi$ as
\ben
\phi =
\frac{1}{\sqrt{2m}}\Psi +
\frac{1}{\sqrt{2m}}\langle\bar{\Psi}^{*}\rangle
\simeq
\frac{1}{\sqrt{2m}}\Psi
\een
on neglecting the anti-particle field fluctuations.  As a result,
zeroes in $\phi$ are also zeroes in $\Phi$.  Further, on neglecting
these fluctuations, $\phi\simeq\pi^{*} /2m$ showing that, in
the non-relativistic limit, $\pi$ is approximately periodic in
imaginary time. Thus the boundary condition $B_{0}$, denoting periodic $\phi$
fields, becomes the boundary condition $b_{0}$ for periodic $\Psi$ fields,
whence we achieve our
goal of showing that, in the non-relativistic limit, the
relativistic partition function
\ben
Z_{\mu}\approx Z_{\mu_{{\rm nr}}} =\int_{b_{0}} {\cal D}\Psi^{*}{\cal D}\Psi
\exp\{-S_{\mu_{\rm nr}}[\Psi^{*},\Psi ]\}\, ,
\label{Zp2}
\een
where $\mu_{{\rm nr}} = \mu - m$ as before. The inclusion of
sources will not change this result, although we shall have to include sources
that couple to
the $\pi^{*}, \pi$ fields.

This demonstration of how calculations with the
relativistic $Z_{\mu}$ of Eq.\ref{Zc} are indistinguishable from those
of the non-relativistic $Z_{\mu_{{\rm nr}}}$ when $\mu\approx m$
can be extended to the case of symmetry breaking, although we shall not be able
to go
beyond the simplest Gaussian approximation in this talk.
Further details are given in our work\cite{altimray2}.

Now that we have demonstrated how to recover the non-relativistic
field theory, we return to the relativistic formalism of
Eq.\ref{S0}, for large and small $\mu$.

\section{\bf A Gaussian Model for Vortex Formation}

Having established the role of the chemical potential in letting us
interpolate between relativistic and non-relativistic regimes in the
initial conditions (with a corresponding interpolation between
relativistic and non-relativistic field theories),
we are now in a position to determine the effect
of these initial conditions on vortex production in a simple model of
Gaussian field fluctuations.

We have already assumed that the initial conditions correspond
to a disordered state in equilibrium for $t < t_{0}$.
The action Eq.\ref{St} was proposed in a context in which the chemical
potential was relegated to the boundary condition.  As we observed
earlier, if we wish to
represent it so as to match the action Eq.\ref{S0}, in which the
chemical potential is manifest, we can accomodate changes in $m$ or
changes in $\mu$ by the requirement that, at
$t = t_{0}$, $m_{0}^{2}(t) = m^{2}(t) - \mu^{2}(t)$ changes sign.
Most simply, we assume that, for $t > t_{0}$, $m_{0}^{2}(t)$
takes the {\it negative} value $m_{0}^{2}(t) = - M^2 <0$ {\it immediately}.
That is,
the potential at the origin has been instantaneously inverted
everywhere, breaking the global $O(2)$ symmetry.   If
$\lambda (t) =  \lambda$ is very {\it weak} then, for times $Mt <
\ln(1/\lambda)$, the $\phi$-field, falling down the hill away from
the metastable vacuum, will not yet have experienced the upturn of
the potential, before the point of inflection.
Thus, for these small times,  $\lambda
(t)$ can also be set to zero, and $p_{t}[\Phi ]$ is Gaussian, as required.

In the relativistic regime $\mu\ll m$, and the $t$-dependence of
$m_{0}$ is entirely carried by $m$, as in Eq.\ref{T/Tc}.  On the other hand,
in the non-relativistic limit, when $\mu\simeq m$ and $m$ is
unchangine in time, $m_{0}^{2}$
changes from
\ben
m_{0}^{2}(t) = m_{0}^{2} = -2m\mu_{0}\, > 0,
\een
say, when $t\leq t_{0}$, to
\ben
m_{0}^{2}(t)  = -M^{2} = -2m\mu_{f}\, < 0,
\label{ms}
\een
when $t > t_0$.
Henceforth, we take $t_{0} = 0$.

The onset of the phase transition at time $t=t_0$ is characterised by
the instabilities of long wavelength fluctuations permitting the growth of
correlations. Although the initial
value of $\langle \phi \rangle$ over any volume is zero, we
anticipate that
the resulting evolution will lead to
domains of constant $\langle \phi \rangle$ phase, whose boundaries will
trap vortices.

Of course, instantaneous change is physically impossible.
Consider small amplitude fluctuations of $\phi_a$, at the top of the
parabolic potential hill.  Long wavelength fluctuations, for which $|{\bf
k}|^2 < M^2$, begin to grow exponentially. If their growth rate
$\Omega (k) = \sqrt{M^2 - |{\bf k}|^2}$ is much slower than
the rate of change of the environment which is causing the quench,
then those long wavelength modes are unable to track the quench.
 It will
turn out that the time-scale at which domains appear in this
instantaneous quench is $t_d = O(M^{-1})$. As long as the time taken
to implement the quench is comparable to $t_d$ and much less than
$t_f = O(M^{-1}ln (1/\lambda )$ the approximation is relevant.
We note that, in the non-relativistic regime, for which $\Omega (k)$ has the
same definition,
\bea
\Omega^2(k) &=& M^2 - |{\bf k}|^2 = 2m\biggl (\mu_{f} -
\frac{k^{2}}{2m}\biggr )
\nonumber
\\
&=& 2m(\mu_{f} - \epsilon (k)).
\eea
Thus the momentum restriction $|{\bf k}|<M$ is just
$\epsilon (k) < \mu_{f}$.

We are now in a position to evaluate $p_t[\Phi]$, or rather $W_{ab}(r;t)$, for
$t > 0$, and
calculate the defect density accordingly.
Before we quote the result we note that
the $i\mu (\phi_{2}\dot{\phi_{1}} - \dot{\phi_{2}}\phi_{1})$
term in $S_{\mu}[\phi ]$ of Eq.\ref{S0} couples the $a =1$ and $a =
2$ fields $\phi_{a}$ together and, in general,
$G_{ab}({\bf
r} -{\bf r}';t,t') = \langle\phi_{a}({\bf r},t)\phi_{a}({\bf
r}',t')\rangle $
is {\it not} diagonal in the $O(2)$ labels.
However, for {\it equal} times diagonal behaviour is restored as
$G_{ab}({\bf r};t,t) = \delta_{ab}G({\bf r};t,t)$ and the
$i\mu (\phi_{2}\dot{\phi_{1}} - \dot{\phi_{2}}\phi_{1})$
term can effectively be discarded.  This leaves us in a situation
for
which Eq.\ref{g1} and Eq.\ref{g2} are satisfied and
the results of Halperin are directly applicable.  That is, $W_{ab}$ is
diagonal,
\ben
W_{ab}(|{\bf r}-{\bf r}'|;t,t) = \delta_{ab}W(|{\bf r}-{\bf r}'|;t,t),
\een
whence $W(|{\bf
r} -{\bf r}'|;t) = \langle\phi_{a}({\bf r},t)\phi_{a}({\bf
r}',t)\rangle $ (no summation), the
thermal Wightman function for either $\phi_{1}$ or $\phi_{2}$.

But for the chemical potential,
this situation of inverted harmonic oscillators was studied many
years ago by Guth and Pi\cite{guth} and Weinberg and Wu\cite{weinberg}.
In the context of domain formation, we refer to
the recent work of Boyanovsky et al.\cite{boyanovsky}, and our
own\cite{alray,altimray}.
For the case in hand, if we make a separation
into the unstable long wavelength modes, for which $|{\bf k}|<M$,
and the short wavelength modes $|{\bf k}|>M$, then  $W(r;t)$
is
\bea
 W(r;t) &=& \int_{|{\bf k}|<M} d \! \! \! / ^3 k
\, e^{i {\bf k} . {\bf x} } C(k;\mu )
\biggl [ 1+ A(k)(\cosh(2\Omega (k)t) - 1 ) \biggr ]
\nonumber
\\
&+& \int_{|{\bf k}|>M} d \! \! \! / ^3 k
\, e^{i {\bf k} . {\bf x} } C(k;\mu )
\biggl [ 1+ a(k)(\cos(2w(k)t) - 1 ) \biggr ]
\label{G}
\eea
with $r = |{\bf x}|$ and
\begin{eqnarray}
\Omega^2(k) &=& M^2 - |{\bf k}|^2,
\nonumber
\\
w^{2}(k) &=& -M^2 + |{\bf k}|^2,
\nonumber
\\
A(k) &=& \frac{1}{2} \biggl (
1+ \frac{\omega ^2(k) }{\Omega^2(k)} \biggr ),
\nonumber
\\
a(k) &=& \frac{1}{2} \biggl (
1- \frac{\omega ^2(k) }{w^2(k)} \biggr ).
\label{defs}
\end{eqnarray}
All the $\mu$-dependence is contained in the factor
\ben
{\cal C}(k, \mu_{0}) = \frac{1}{2\omega (k)}\biggl [\frac{e^{-\beta_{0}\omega}
- -e^{\beta_{0}\omega}}
{e^{\beta_{0} (m + \mu_{0})}  - e^{-\beta_{0}\omega} - e^{\beta_{0}\omega}
+ e^{-\beta_{0} (m + \mu_{0})}}
\biggr ].
\een
where
\ben
\omega ^2(k) = |{\bf k}|^2 + m_0^2.
\een
In the zero chemical potential limit $\mu = m_{0} + \mu_{0} = 0$, ${\cal C}(k,
\mu_{0})$ takes the familiar form
\ben
{\cal C}(k, \mu_{0}) = \frac{1}{2\omega (k)}coth(\beta_{0}\omega (k)/2)
\label{cr}
\een
On the other hand, in the non-relativistic limit $\epsilon (k) = {\bf k}^{2}/2m
\simeq \omega
- - m$ and $\mu_{0}\ll m$ ${\cal C}(k, \mu_{0})$ is the equally
familiar Bose distribution
\ben
{\cal C}(k, \mu_{0}) \simeq  \frac{1}{2\omega (k)}\bigg (\frac{1}{1 -
e^{-\beta_{0}(\epsilon
(k) - \mu_{0})}}\Bigg ).
\label{C}
\een
Further, in the high temperature relativistic limit $T_{0}\gg m_{0}$,
${\cal C}(k, \mu_{0})$ of Eq.\ref{cr} simplifies as
\ben
\frac{1}{2\omega (k)}coth(\beta_{0}\omega (k)/2)
\simeq \frac{T_{0}}{{\bf k}^{2} + m_{0}^{2}}.
\label{wc}
\een
Equally, in the non-relativistic limit when $\epsilon (k) - \mu_{0} \ll
T_{0}$,
\bea
{\cal C}(k, \mu_{0})
&\simeq &
 \frac{T_{0}}{\epsilon (k) + |\mu_{0}|}
\label{C2}
\nonumber
\\
&\propto &\frac{T_{0}}{{\bf k}^{2} + m_{0}^{2}}
\label{wC}
\eea
on using the definition of $m_{0}$ given earler in Eq.\ref{ms}.
{}From Eq.\ref{ni} onwards it follows that the overall scale of $W$ is
immaterial to the vortex density.
As a result, in these regimes the relativistic and non-relativistic
$W(r,t)$ are
{\it identical},
once we take the identifications of Eq.\ref{ms} into account.

We observe that, if we were to use $W(r;t)$ of Eq.\ref{G} as it
stands, then both $W(0;t)$ and $W''(0;t)$ necessarily suffer from
ultraviolet divergences.
However, the string thickness at the end of the quench will be $O(M^{-1})$.
It is the zeroes  coarse-grained to this scale that will provide the
subsequent network.  Thus, if we do not probe the field zeroes
within a string we need consider only the first term
in Eq.\ref{G}.  There is another point. Even before the quench begins
there is a high density of line zeroes coarse-grained to this same
scale $O(M^{-1})$ in the initial equilibrium phase.
However, these modes are entirely transient.
If we were to calculate the correlations of $\rho_{i}(\bf{x})$ at
{\it different} times $t$ and $t'$, we would find rapidly
oscillating behaviuor with period $\Delta t = O(m^{-1})$.  On the
other hand, a calculaation of the density correlations at different
times from the unstable modes in Eq.\ref{G} does not give
oscillatory, but damped, behaviour.  It is the residue of the
strings produced by the unstable modes that survives to produce the
network, and the transient strings can be ignored.  Henceforth, we
retain only the first term
\ben
 W(r;t) = \int_{|{\bf k}|<M} d \! \! \! / ^3 k
\, e^{i {\bf k} . {\bf x} }{\cal C}(k, \mu_{0})
\biggl [ 1+ A(k)(\cosh(2\Omega (k)t) - 1 ) \biggr ]
\label{Gl}
\een
of Eq.\ref{G}.  In the two critical regimes where Eq.\ref{wc} and
Eq.\ref{wC} are valid, $W(r;t)$ is again the same for both cases.

Even though the approximation is only
valid for small times, there is a regime $Mt\ge 1$, for small couplings, in
which
$t$ is large enough for $\cosh(2Mt) \approx \frac{1}{2} \exp(2M
t)$ and yet $Mt$ is still smaller than the time $O(\ln 1/\lambda)$ at
which the fluctuations begin to sample the ground-state manifold.
In this regime
\ben
W(r;t)\simeq \int_{|{\bf k}|<M} d \! \! \! / ^3 k
\, {\cal C}(k, \mu)
e^{i {\bf k} . {\bf x} }
A(k)\;e^{2\Omega (k)t}.
\een
In
these circumstances the integral at time t is dominated by a
peak in the integrand $k^{2} e^{2\Omega (k)t}$ at $k$ around $k_c$, where
\ben
t k_c^2 = M \biggl ( 1 + O \biggl ( \frac {1}{Mt}
\biggr ) \biggr ),
\een
and we have assumed $M$ and $m_{0}$ to be comparable.
The effect of changing $\beta_0$ is only visible in the $O(1/Mt)$
term.
In fact, we are being unnecessarily restrictive in wanting to
preserve the identical behaviour of Eq.\ref{wc} and Eq.\ref{wC}.
All that is required for
identical {\it leading} behaviour in relativistic and non-relativistic
regimes is that ${\cal C}(k, \mu_{0})$ varies
slowly in the vicinity of the peak of the integrand at
$tk_{c}^{2}\simeq M$.  [$A(k)$ is already slowly varying].
This is the case when, allowing for
coefficients $O(1)$,
\ben
1 < \frac{\mu_{f}}{|\mu_{0}|} = \frac{M^{2}}{m_{0}^{2}}\ll tM <
\ln\biggl (\frac{1}{\lambda}\biggr )
\label{lims}
\een
where, in the same spirit, we have taken $|\mu_{0}| < \mu_{f}$.  The
upper bound on $tM$ is a reminder that interactions are always
present, and the Gaussian approximation must fail as soon as the
field fluctuations have extended to the true ground-states at the
minima of the potential.  The lower bound is necessary for the
integrand to be peaked strongly so that the saddle-point
approximation is valid.  As long as Eq.\ref{lims} is basically
correct, any difference between the relativistic and
non-relativistic regimes will be non-leading.

Assuming these limits, we recover what would have been our first naive guess
for a correlation
function based on the growth of the unstable modes,
\ben
W(r;t)\simeq\int_{|{\bf k}|<M} d \! \! \! / ^3 k
\, e^{i {\bf k} . {\bf r} }
\;e^{2\Omega (k)t},
\label{Wapp}
\een
We understand the dominance of wavevectors at $k_{c}^{2} = M/t$ in
the integrand as indicating the formation of domains of mean size
\ben
\xi (t) = O(\sqrt{t/M})
\label{xit}
\een
 once $Mt > 1$.  Specifically, we take
$\xi (t) = 2 \sqrt{t/M}$.
Once $Mt > 1$ then $\xi (t) >
M^{-1}$, where $M^{-1}$ characterises the cold vortex radius. In the weak
coupling
approximation
individual domains become
large enough to accomodate many vortices before the approximation
breaks down.  There is no difficulty with
causality since domains increase in size as $\dot{\xi} =
\frac{1}{\sqrt{Mt}} < 1$.
On neglecting terms exponentially small in $Mt$, $W(r;t)$ of
Eq.\ref{Wapp} can be further rewritten as
\bea
 W(r;t ) & \simeq &
 e^{2Mt}\int_{0}^{\infty} dk\,\, \sinc(kr)\, k^{2} e^{-tk^{2}/M}
\label{WG}
\\
&=& W(0;t)\,exp\biggl (-\frac{r^{2}}{\xi^{2}(t)}\biggr )
\label{WG2}
\eea
where
\ben
W(0;t) \approx C \frac {e^{2M t} } {(M t) ^{3/2} },
\een
for some C. The exponential growth
of $W(0;t)$ in $t$ reflects the way the field amplitudes fall
off the hill centred at $\Phi = 0$.  With the peaking in wavelength $l
= k^{-1}$
understood as indicating the appearence of domains of characteristic
linear dimension $\xi (t)$, the Gaussian in $r$ is a reflection
of the rms variation $\Delta\xi$ in domain size $\xi$.
This variation is large.
If we isolate the Gaussian saddle-point in Eq.\ref{Wapp} as
\ben
W(r;t)\simeq  e^{2Mt}\int_{0}^{M} dk\,\, \sinc(kr)\, k_{c}^{2} e^{-(k -
k_{c})^{2}/2(\Delta k)^{2}}
\label{Wapp2a}
\een
 then
\ben
\frac{\Delta\xi}{\xi} = \frac{\Delta k}{k_{c}} = \frac{1}{2}.
\label{rms}
\een

To calculate the number density of vortices at early times we
insert the expression Eq.\ref{WG} for $W$ into the equations derived
earlier, to find
\ben
n(t) = \frac{1}{\pi}
\frac{1}{\xi(t)^2}.
\label{nt}
\een
We note that the dependence on time
$t$ of both the density and density correlations
 is only through the correlation length $\xi (t)$.  We have a {\it
scaling} solution in which, as the
domains of coherent field form and expand, the interstring distance
grows accordingly.  Since the only way the defect density can
decrease without the background space-time expanding is by
defect-antidefect annihilation, we deduce that the coalescence of
domains proceeds by the annihilation of small loops of
string.  However, because the density of vortices only depends on $\xi
(t)$ in this early stage, the fraction of string in `infinite'
string remains constant.
Thus, at the same time as small loops disappear,
other loops must rearrange themselves so that the length of
`infinite' string decreases accordingly.
Finally, there is roughly
one string zero per coherence area,  a long held belief for whatever
mechanism.

There is one final concern.  A necessary condition for this rolling
down of the field to be valid is that the initial field fluctuations  should
be small enough that there is no significant probability that the
field is already in the true vacuum. As long as we do not begin the
quench from too close to the transition, there is, in fact, no
difficulty.  See our work\cite{altimray} for more details.

\subsection{Vortex Density Correlations}

In addition to the gross vortex density Eq.\ref{nt} we can calculate
the density-density correlation functions $C_{ij}(r;t)$ of
Eq.\ref{ddc}, identical for both the relativistic  plasma and the
non-relativistic medium.

Yet again, as in the case of the density $n(t)$, the $t$-dependence  of
$C_{ij}$ only occurs implicitly through $\xi (t)$.
The simple analytic form of Eq.\ref{WG} enables us to calculate $A$ and
$B$, up to exponentially small terms in $Mt$, as
\bea
A(r;t) &=& \frac{2}{\pi^{2}\xi^{4}(t)}
\frac{e^{-2r^{2}/\xi^{2}(t)}}{(1 - e^{-2r^{2}/\xi^{2}(t)})^{2}}
\biggl [(1 - e^{-2r^{2}/\xi^{2}(t)}) - 2\frac{r^{2}}{\xi^{2}(t)}
\biggr ] < 0,
\nonumber
\\
B(r;t) &=& \frac{2}{\pi^{2}\xi^{4}(t)}
\frac{e^{-2r^{2}/\xi^{2}(t)}}{(1 - e^{-2r^{2}/\xi^{2}(t)})} > 0,
\label{corr1}
\eea
Suppose that ${\bf r} = (0,0,r)$.  Then
\bea
C_{11} = C_{22} &=& A(r;t) -   B(r;t)
\nonumber
\\
&=& -\frac{2}{\pi^{2}\xi^{4}(t)}
\biggl (\frac{2r^{2}}{\xi^{2}(t)}\biggr )
\frac{e^{-2r^{2}/\xi^{2}(t)}}{(1 - e^{-2r^{2}/\xi^{2}(t)})^{2}} < 0.
\label{corr2}
\eea
That is, we have {\it anticorrelation} of densities for parallel
directions (and positive $B$ for orthogonal ones).

It  is useful to expand Eq.\ref{corr1} and Eq.\ref{corr2} for $r <
\xi$.  On normalising by a factor of $n^2$,
\bea
\frac{A(r;t)}{n^{2}(t)} &=& -1 + O\biggl (\frac{r^{2}}{\xi^{2}}\biggr ) < 0,
\nonumber
\\
\frac{B(r;t)}{n^{2}(t)} &=& \biggl (\frac{\xi^{2}(t)}{r^{2}}\biggr )
+ O(1) > 0.
\label{corr3}
\eea
For $r > \xi$  there is exponential falloff but, from Eq.\ref{corr3}
we see that, in units of $n(t)^{-2}$, the
anticorrelation is large.
Since strings with a
long persistence length would imply {\it positive} parallel
correlations, we can
interpret these anticorrelations (and others) as a reflection of an increased
string bendiness.
Although it is difficult to be precise, this suggests a significant
amount of string in small loops.
This is an important issue since,
as we noted, early universe
cosmology requires infinite string if string is to be the source
of large-scale structure.  In practice, some is enough.

There  is another, indirect, way in which we can see the tendency for
more string to
be in small loops than we might have thought.
In numerical simulations of string networks,
the rule of thumb for a {\it regular} domain structure in field
phase is that a
significant fraction of string, if not most, is
infinite string.

However, as we saw earlier in Eq.\ref{rms}, we do not have a regular
domain structure in our model but have domains with a large variance
$\Delta\xi /\xi = \frac{1}{2}$.
Unfortunately, the domain structure that we have here is not yet appropriate
for a direct comparison since, as well as the domain size, the field magnitude
has a variance about $\langle |\phi|\rangle = O(M\,e^{Mt})$.  The
work of Guth and Pi\cite{guth} shows that this can be parametrised
by an effective dispersion in $t$ in $\langle |\phi|\rangle$ of
$\Delta t = O(M^{-1})$.  Nonetheless, consider an
idealised case in which domain growth stops instantaneously because
of back-reaction at some time $t_{f}$.  The distribution of strings will then
be as
above for $\xi = \xi (t_{f})$, while the field adjusts to the vacuum manifold,
without
changing phase, in each domain.  We would expect some
string-antistring annihilation to continue while this adjustment
occurs, so that the $n(t)$ calculated previously is an overestimate
of the string density at the end of the transition.  However, the
domains will still be of varying size with variance still given
approximately by Eq.\ref{rms}.  The inclusion of domain variance in
numerical simulation of string networks (albeit in a different way
from that proposed here) shows\cite{andy} that, the greater the
variance, the more string is in small loops.
Beyond observing that there
seems to be some infinite string, we will say no more.

All our results are for quenches that go from significantly above the
transition to significantly below it.  These give the {\it smallest} value
of $\Delta\xi /\xi$ possible and, plausibly, the best chance of
producing infinite string. Had we begun closer to the transition,
where the initial fluctuations are larger, the dispersion in domain
size, and consequently the fraction of string in small loops, would
be even greater.

\section{\bf Conclusions}

In this paper we have shown how global O(2) vortices appear, at a
quench from the ordered to disordered state, as a consequence of the
growth of unstable Gaussian long wavelength fluctuations.  Most
importantly, in the light of discussions about the extent to which
vortex production in low-temperature many-body systems simulates
vortex production in the early universe, our model supports the
analogy.  Specifically, with our simple assumptions, vortex
production is {\it identical} in both a relativistic
high-temperature quench, in which the initial state is characterised
by $T\gg m$, and in a non-relativistic density quench in which the
initial state is described by $T\ll m$ (to freeze out antiparticles)
with a chemical potential $\mu_{{\rm nr}}\simeq T$.  All  that is
required is an appropriate translation of the parameters from the
one case to the other.  To see this, we have shown how to isolate
the non-relativistic particle/antiparticle sectors of the
relativistic path-integrals that we use in our calculations. The
$O(2)$ symmetry, whose breaking leads to the production of vortices,
leads to the conservation of particle number minus antiparticle
number in the former case, and to the conservation of particle
number in the latter.

In our simple model of Gaussian fluctuations the resulting string
configurations scale as a function of the correlation length $\xi
(t) = O(t^{\frac{1}{2}})$, at about one vortex/correlation area.
This is compatible with the Kibble mechanism for vortex production
on domain boundaries upon phase separation.
For a weak coupling theory the domain cross-sections are
significantly larger than a vortex cross-section at the largest
times for which the approximations are valid.  However,  there is a large
variance in their size, with $\Delta\xi /\xi = \frac{1}{2}$.
Because of this there is more string in
small loops than we might have anticipated.

We stress that our model can, at best, describe weak coupling
systems for the short times while the domains are growing
 before the defects freeze out.  This is
unsatisfactory for most early universe applications and for
low-temperature many-body systems.  However, we know in
principle\cite{boyanovsky}
how to include back-reaction (still within the
context of a Gaussian approximation) to slow down domain growth as
the field fluctuations spread to the ground-state manifold.  The
identity of the weak-coupling results of the two regimes should
survive to this case also, although it will probably lead to different
conclusions from those above.  Further, there is no difficulty in
principle of embedding these results in a FRW metric, along the
lines of\cite{hector}.  This is being actively pursued.


\begin{thebibliography}{99}

\bibitem{kibble1} T.W.B. Kibble, {\it J. Phys.} {\bf A9} (1976) 1387.

\bibitem{shellard} E.P.S Shellard and A. Vilenkin, {\it Cosmic Strings
and other Topological Defects} (Cambridge University Press, 1994)


\bibitem{lancaster} P.C. Hendry, N.S. Lawson, R.A.M. Lee, P.V.E. McClintock
and C.D.H. Williams, {\it Nature} {\bf 368} (1994) 314.

\bibitem{helsinki} V.M.H. Ruutu, M. Krusius, U. Parts,
G.E. Volovik,  {\it Neutron Mediated Vortex Nucleation in
Rotating Superfluid $^{3}He-B$}, (Low Temperature Laboratory, Helsinki
University of
Technology, 02150 Espoo, Finland), in preparation.

\bibitem{lancaster2} P.C. Hendry, N.S. Lawson, R.A.M. Lee, P.V.E. McClintock
and C.D.H. Williams, to be published in the Proceedings of the NATO
Advanced Study Institute on {\it Topological Defects}, Newton Institute,
Cambridge
1994, (Plenum Press, to appear).


\bibitem{zurek1} W.H. Zurek, {\it Nature} {\bf 317} (1985) 505,
{\it Acta Physica Polonica} {\bf B24} (1993) 1301.

\bibitem{zurek2} W.H. Zurek, to be published in the Proceedings of the NATO
Advanced Study Institute on {\it Topological Defects}, Newton Institute,
Cambridge
1994, (Plenum Press, to appear).

\bibitem{alray} A.J. Gill and R.J. Rivers,
{\it Phys. Rev.}{\bf D51} (1995) 6949.

\bibitem{timray} R.J. Rivers and T.S. Evans, {\it The Production of
Strings and Vortices at Phase Transitions},
to be published in the Proceedings of the NATO
Advanced Study Institute on {\it Topological Defects}, Newton Institute,
Cambridge
1994, (Plenum Press, to appear).

\bibitem{altimray} A.J. Gill, T.S. Evans and R.J. Rivers, {\it
Vortex Production in Non-Relativistic and Relativistic Media},
Imperial College Preprint, Imperial/TP/94-95/48, July 1995.

\bibitem{altimray2} A.J. Gill, T.S. Evans and R.J. Rivers, {\it
The Chemical Potential and the Non-Relativistic Limit},
Imperial College Preprint (in preparation).

\bibitem{halperin} B.I. Halperin, {\it Statistical Mechanics of
Topological Defects}, published in {\it Physics of Defects},
proceedings of Les Houches, Session XXXV 1980 NATO
ASI, editors Balian, Kl\'{e}man and Poirier (North-Holland Press, 1981) p.816.

\bibitem{schwinger} J. Schwinger, {\it J. Math. Phys.} {\bf 2}
(1961) 407.

\bibitem{mahantapa} K.T. Mahanthappa and P.M. Bakshi, {\it J.M. Phys.} {\bf
4} (1963) 1; ibid, (1963) 12.

\bibitem{keldysh} L.V. Keldysh, {\it Sov. Phys. JETP} {\bf 20} (1965) 1018.

\bibitem{Go4}  K-C. Chou, Z-B. Su, B-L. Hao and L. Yu, Phys.
Rep. {\bf 118} (1985) 1.

\bibitem{boyanovsky} D. Boyanovsky, Da-Shin Lee and Anupam Singh,
{\it Phys. Rev.} {\bf D48} (1993) 800.

\bibitem{guth} A. Guth and S-Y. Pi, {\it Phys. Rev.} {\bf D32} (1985) 1899.

\bibitem{weinberg} E.J. Weinberg and A. Wu, {\it Phys. Rev.} {\bf D36} (1987)
2474.


\bibitem{kapusta} J.I. Kapusta, {\it Finite Temperature Field
Theory}, (Cambridge University Press, 1989).
\\
J.I. Kapusta, {\it Phys. Rev.} {\bf D24} (1981) 426.

\bibitem{art} H.E. Haber and H.A. Weldon, {\it Phys. Rev.} {\bf D25} (1982)
502.

\bibitem{dodelson} J. Bernstein and S. Dodelson, {\it Phys. Rev. Lett.}
{\bf 66} (1991) 683.
\\
K.M. Benson, J. Bernstein and S. Dodelson, {\it Phys. Rev.}
{\bf D44} (1991) 2480.

\bibitem{andy} A. Yates and T.W.B. Kibble, Imperial College preprint
Imperial/TP/94-95/49.

\bibitem{hector} D. Boyanovsky, H.J. de Vega, R. Holman, {\it Phys. Rev.}
{\bf D49} (1994) 2769.

\end{thebibliography}
\end{document}